# On the nature of inertial mass

**Alexander L. Dmitriev**

St-Petersburg State University of Information Technologies, Mechanics and Optics
49, Kronverksky Prospect, St-Petersburg, 197101, Russia

E-mail: dalexl@rol.ru

**Abstract**
It is shown that gravitational nature of inertial mass (Mach's principle) agrees with idea of interaction of gravitational and electromagnetic forces and does not contradict the laws of classical mechanics. According to the simple phenomenological model the body inertial mass is directly proportional to its gravitational mass and the sum of coefficients $\alpha_p$ and $\alpha_c$ which characterize degrees of interaction of gravitational forces in accelerated motion of the body in accompanying and opposite directions relative to the gravitational force.

PACS number: 04.80. Cc

Inertial mass of the body which is a component of the second law of dynamics has long been an object of scientific discussions. A wide-known concept indicating the cause and physical sense of inertial mass is presented by Mach's principle according to which the inertial mass of the body is a result of that body gravitational interaction with all surrounding bodies of the universe [1,2]. Gravitational interaction of bodies is directly connected with the concept of gravitation propagation rate. Extreme remoteness of massive formations – stars, galaxies, clusters of galaxies, etc. – provides grounds for assumption that the condition for Mach's principle realization must be rather high speed of gravitation propagation which might considerably exceed light speed.

The present paper shows that the gravitational nature of inertial mass when described as a phenomenon naturally agrees with the concept of interaction of electromagnetic and gravitational forces. Such an interaction is characterized by dependence of gravity force acceleration change $\Delta\vec{g}$ and acceleration $\vec{a}$ which is caused by action of external, for example, electromagnetic forces on the test body [3]. Similar to Lenz's law, directions of vectors $\Delta\vec{g}$ и $\vec{a}$ are mutually opposite, and coefficients of proportionality $\alpha_p, \alpha_c$ depend on acceleration direction $\vec{a}$ relative to gravitational force acceleration $\vec{g}_0$. In the first (linear) approximation, "the induction" of gravity force acceleration is equal to

$$\Delta\vec{g}_{p,c} = -\alpha_{p,c}\vec{a}, \qquad (1)$$

where $\alpha_p$ corresponds to parallel vectors $\vec{a}$ and $\vec{g}_0$, and $\alpha_c$ - to antiparallel ones – Fig. 1.

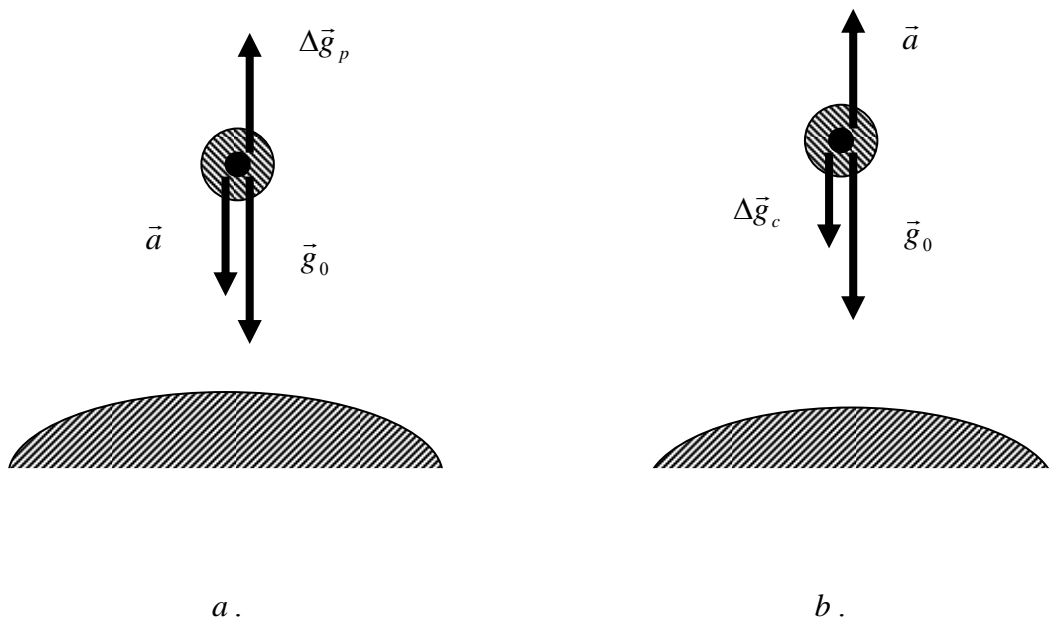

*a*.  *b*.

Fig. 1. a. Changes in gravity force acceleration acting on test body while body is falling down with acceleration.  b. Changes in gravity force acceleration while body is moving up with acceleration.

Validity of equation (1) is confirmed by experiments on precise weighing of bodies moving with acceleration, as well as - indirectly – by negative temperature dependence of weights of bodies [4-6]. Absolute magnitude of interaction coefficients $\alpha_{p,c}$ measured in experiments with non-magnetic metal samples is comparatively high - $\alpha_c \sim 10^{-2}$, and the order of positive difference magnitude - $(\alpha_p - \alpha_c) \sim 10^{-7}$.

Let's assume that in conditional-inertial system of coordinates related to "infinitely remote" masses, a test body with gravitational mass $m_g$ is stationary in the initial state (Fig. 2a).

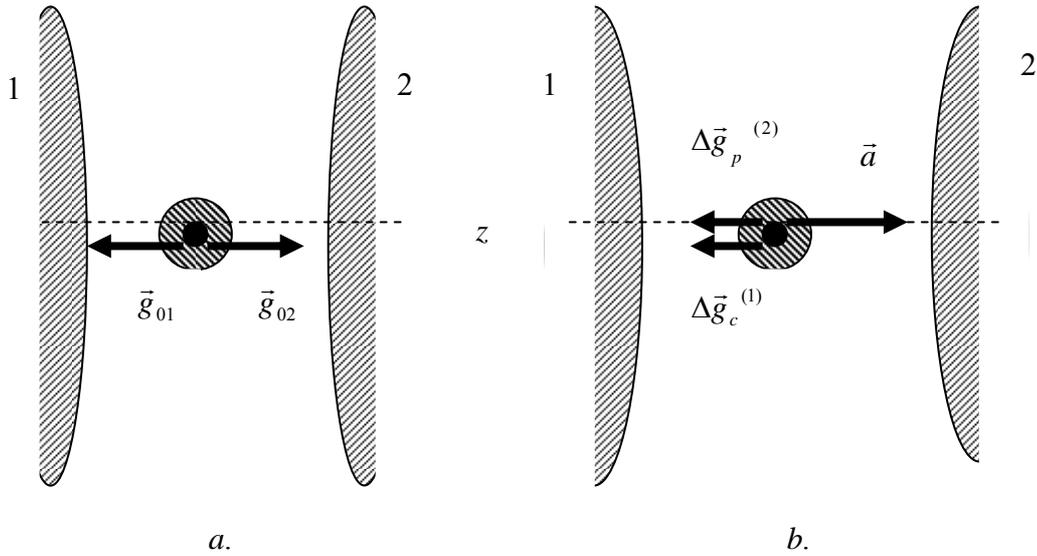

Fig. 2. $a$. Test body in counterbalanced state. $b$. Increments in gravity force acceleration under effect of outside non-gravitational force.

In isotropic space along a randomly chosen axis $z$, the resultant force of gravitational forces acting on the body from the side of remote masses 1, 2 located on different sides of the body is equal to zero. In so doing, the accelerations corresponding to those forces which are the sum of projections on axis $z$ of gravity forces accelerations created by such masses are equal in magnitude $|\vec{g}_{01}| = |\vec{g}_{02}|$. If there is an outside non-gravitational force $\vec{F}$ acting on the body along the indicated direction, then according to the second law of dynamics the corresponding acceleration $\vec{a}$ of the body is equal to

$$\vec{a} = \vec{F}/m_i, \qquad (2)$$

where $m_i$ - body inertial mass. Such an acceleration, according to (1), causes changes $\Delta\vec{g}_c^{(1)}$ and $\Delta\vec{g}_p^{(2)}$ of gravity force accelerations acting on the test body from the side of remote masses 1, 2, with directions of vectors $\Delta\vec{g}_c^{(1)}$ and $\Delta\vec{g}_p^{(2)}$ coinciding (Fig. 2b).

According to the third law of dynamics the resultant force of those applied to the body is equal to zero,

$$m_g(\vec{g}_{01} + \Delta\vec{g}_c^{(1)}) + m_g(\vec{g}_{02} + \Delta\vec{g}_p^{(2)}) + m_i\vec{a} = 0. \qquad (3)$$

Whence the inertial mass of test body

$$m_i = (\alpha_p + \alpha_c)m_g \qquad (4)$$

Expression (4) explains the physical sense of inertial mass: body inertial mass is proportional to its gravitational mass, caused by interaction of gravitational and electromagnetic forces applied to the body, and determined by magnitudes of coefficients $\alpha_p$, $\alpha_c$ which characterize the degree of such an interaction. If the interaction of gravitational and electromagnetic forces were nonexistent, that is with $\alpha_c = \alpha_p = 0$, the body inertial mass would be equal to zero as well. Such an assumption directly agrees with Mach principle according to which inertia of matter is determined by masses surrounding it. On the other hand, proportionality of gravitational and inertial masses is proved by numerous experiments and agrees with corpuscular model of matter: an increase in number of particles making up test body proportionally increases both inertial and gravitational masses. Mass ratio $m_i/m_g$ is determined by magnitudes of coefficients $\alpha_p$, $\alpha_c$ which might be constants; this expression can be equated to one by choosing corresponding magnitude of gravitational constant.

Thus, the gravitational nature of inertial mass does not contradict the principles of classical mechanics supplemented by concept of interaction of gravitational and electromagnetic forces (in particular, elastic forces). Experimental and theoretical investigations of gravity force acceleration "induction" caused by accelerated motion of the body under effect of outside non-gravitational forces will contribute to development of physics of gravitation and its supplements.